\documentstyle[12pt,epsfig]{article}

\newcommand{\be}{\begin{equation}}
\newcommand{\ee}{\end{equation}}
\newcommand{\ba}{\begin{eqnarray}}
\newcommand{\ea}{\end{eqnarray}}

\begin{document}

\begin{flushright}
{\bf FIAN/TD-04/2000}\\
{\bf BNL-NT-00/12}\\
\end{flushright}

\bigskip

\bigskip

\begin{center}
{\bf ON TRANSVERSE ENERGY PRODUCTION IN HADRON COLLISIONS}
\end{center}

\medskip

\begin{center}
\bf Andrei Leonidov$^{(a),(b)}$
\end{center}

\medskip

\begin{center}
{\it $(a)$ Physics Department, Brookhaven National Laboratory\\
Upton, NY 11973, USA}
\end{center}
 
\begin{center}
{\it $(b)$ Theoretical Physics Department, P.N.~Lebedev Physics Institute \\
117924 Leninsky pr. 53, Moscow, Russia \footnote{Permanent address}}
\end{center}

\medskip

\begin{center}
{\bf Abstract}
\end{center}

The transverse energy    spectrum in the  unit rapidity window in
$p {\bar p}$ collisions at $\sqrt{s}=540\, {\rm GeV}$ is calculated to
the next-to-leading order accuracy $O(\alpha_s^3)$ and compared to the
experimental data by UA(2) collaboration \cite{UA2}.
We show that the calculated spectrum starts matching experimental data
only at relatively large transverse energy  $E_{\perp} \sim 60\,{\rm GeV}$
and is in essential disagreement with it both in shape and magnitude
at lower transverse energies. The data are well reproduced by HIJING
Monte-Carlo generator indicating the crucial importance of all-order
effects in perturbation theory as well as those of hadronization
in describing the transverse energy production in hadron collisions
at small and intermediate transverse energies.

\newpage

 In this note we calculate to the next-to-leading (NLO) order accuracy
the transverse energy spectrum in the central  rapidity
window within perturbative QCD and compare it to experimental data
obtained by UA2 collaboration \cite{UA2}.  The NLO calculation of
a generic jet cross section requires using a
so-called jet defining algorithm specifying the resolution for the jet
to be observed, for example,  the angular size of the jet-defining  cone,
see e.g. \cite{S},
The cross section in question is calculated by integrating the differential
one  over the phase space, with the integration domain restricted by the
jet characteristics  fixed by the jet-defining alogorithm.
Schematically the NLO distribution of the transverse energy produced into a
given rapidity interval $y_a < y < y_b$ is given, to the $O(\alpha_s^3)$
order,  by
\begin{eqnarray}
\frac{d\sigma}{dE_\perp}=\int D^2PS  \frac{d\sigma}{d^4p_1d^4p_2}
\delta(E_\perp-\sum\limits_{i=1}^2 |p_{\perp
i}|\theta(y_{min}<y_i<y_{max}))\nonumber\\
+\int D^3PS  \frac{d\sigma}{d^4p_1d^4p_2d^4p_3}
\delta(E_\perp-\sum\limits_{i=1}^3 |p_{\perp i}|\theta(y_{min}<y_i<y_{max}))
\end{eqnarray}
 where the first contribution corresponds to the two-particle
final state and the second contribution  to the three-particle one. The
two-particle contribution has to be computed with one-loop corrections taken
into account.

 In perturbative QCD one can rigorously compute only infrared safe
quantities \cite{S}, in which the divergences originating from real and
virtual gluon contributions cancel each other, so that adding very soft gluon
does not change the answer. It is easy to convince oneself, that the
transverse energy distribution into a given rapidity interval Eq.~(1)
satisfies the above requirement\footnote{For a formal definition 
of infrare safety see, e.g., \cite{KS}}.

 The calculation  of  transverse energy spectrum in $p {\bar p}$ collisions
was  performed in \cite{LO1} using the Monte-Carlo code developed by
Kunzst and Soper \cite{KS}, and a  "jet" definition appropriate for transverse
energy production Eq.~(1).   Recently the
transverse energy production to the NLO accuracy was also discussed in 
\cite{ET}, where the only the moments of transverse energy distribution
were discussed and no comparison with experimentally observed transverse
energy spectra was made.

\begin{figure}[h]
 \begin{center}
 \epsfig{file=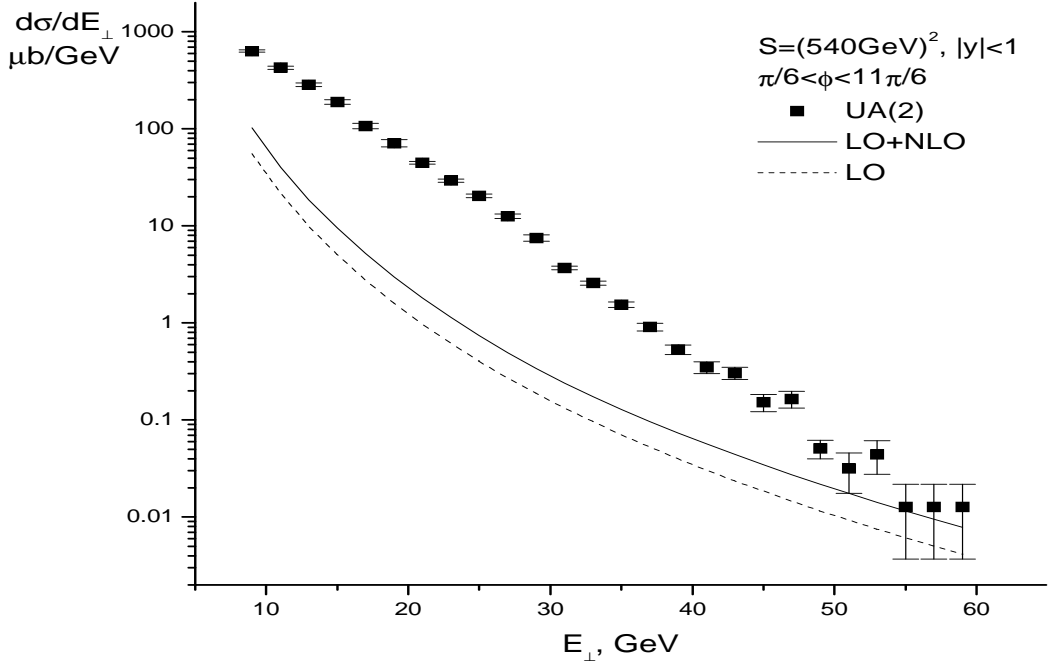,height=9cm,width=14cm}
 \end{center}
 \caption{Transverse energy spectrum in $p {\bar p}$ collisions calculated 
  within LO+NLO accuracy in perturbative QCD vs the experimental data by
  UA2 collaboration \cite{UA2}}
 \label{fua2}
\end{figure}

In Fig.~\ref{fua2} we compare the  LO and LO+NLO transverse energy spectra for
$p{\bar p}$ collisions, calculated as in \cite{LO1}, with the experimental
data on transverse energy distribution in the central rapidity window
$|y| < 1$ and 
asimuthal coverage $\pi/6 \le \varphi \le 11 \pi/6$ at $\sqrt{s}=540$ GeV
measured by UA2 Collaboration \cite{UA2}.
We see that the perturbative LO+NLO calculations start merging with the
experimental data only around quite a large scale  $E_{\perp} \sim 60$\ GeV.
It is interesting to note, that it is precisely around this energy, that the
space of experimental events starts to be dominated by two-jet configurations
\cite{UA2}. This means that only starting from these transverse energies the 
assumptions behind the perturbative calculation (collinear factorization at 
leading twist, explicit account for all contributions of a given order in
$\alpha_s$) are becoming adequate to the observed physical
process of  transverse energy production providing the required duality
between the  description of  dominant configuration contributing to
transverse energy  production at this order in perturbation heory
and the final state transverse energy carried by hadrons. 
At $E_{\perp} \le 50$\ GeV the calculated spectrum is in radical disagreement
with the experimental one both in shape and magnitude 
calculated and observed spectrum is very large indicating the inadequacy 
of the considered $O(\alpha_s^3)$  perturbative calculation
in this domain. Let us mention here, that it is currently
impossible to improve the results of the above calculation, because
neither  calculations of higher order nor infinite order ressummation
for this process are currently available
\footnote{For discussion of the contribution of the initial and 
final state radiation see \cite{infin}}.

\begin{figure}[h]
 \begin{center}
 \epsfig{file=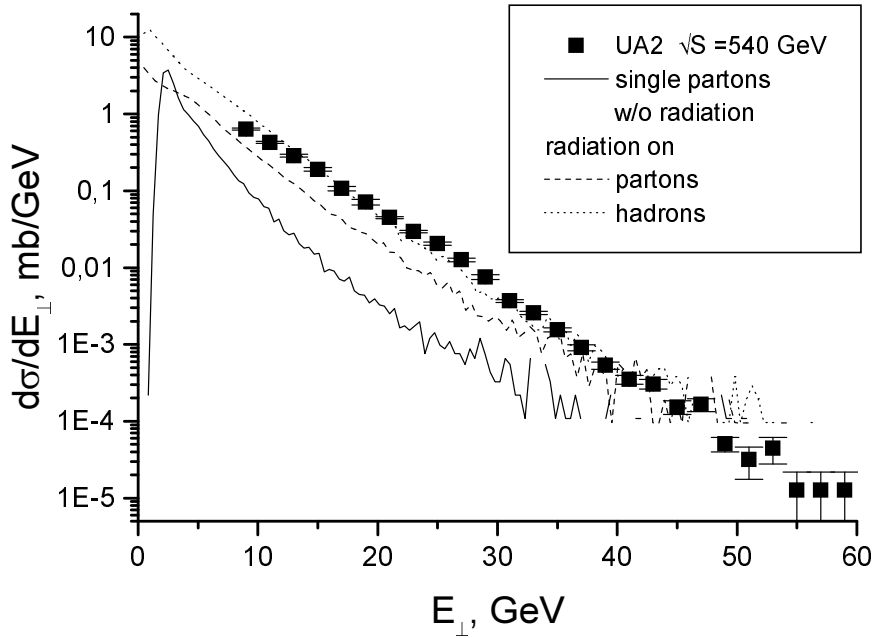,height=9cm,width=12cm}
 \end{center}
 \caption{Transverse energy spectrum in $p {\bar p}$ collisions 
 calculated in HIJING vs the experimental data by UA2 collaboration
 \cite{UA2}}
 \label{fua2hj}
\end{figure}

In practical terms this means that additional model assumptions are needed
to achieve agreement with experimental data  strongly indicating 
that higher order corrections and higher twist effects have to be 
taken into account (in a necessarily model-dependent way) in order to describe
them. In the popular Monte-Carlo generators such as PYTHIA \cite{PYTHIA}
and HIJING \cite{HIJING} such effects as multiple binary parton-parton 
collisions, initial and final state radiation and transverse energy
production during  hadronization are included. In Fig.~\ref{fua2hj} we
compare the same experimental data by UA(2) \cite{UA2} with the spectrum
calcuated with HIJING event generator. To show the relative importance 
of different dynamical mechanims, in Fig.~\ref{fua2hj} we plot the
contributions from the hard parton scattering without initial and final
state radiation, full partonic contribution and, finally, the transverse
energy spectrum of final hadrons.    
We see, that taking into account additional partonic sources
such as, e.g.,  initial and final state radiation, allows to reproduce
the (exponential) form of the spectrum, but still not the magnitude.
The remaining gap is filled in
by soft contributions due to transverse energy production from 
decaying stretched hadronic strings. The results confirm those
earlier communicated to the author by M.~Gyulassy \cite{MG}.
 Let us note, that the
spectrum calculated in  HIJING is somewhat steeper than the experimental
one. Additional fine-tuning can be achieved by probing different
structure functions.

The above results clearly demonstrate that in order to reproduce the
experimentally observed transverse energy spectrum, one has to account
for complicated mechanisms of parton production, such as initial and final
state radiation accompanying hard parton-parton scattering, production of
gluonic kinks by strings, as well as for nonperturbative transverse energy
production at hadronization stage. This statement is a calorimetric
analog of of the well-known importance of the minijet component in 
describing the tails of the multiplicity distributions, \cite{SZ}
and \cite{HIJING} (see, however, the alternative explanation 
based on accounting for multipomeron contributions described in \cite{MW}).
 
Let us note that the result has straightforward implications for
describing the early stages of heavy ion collisions. In most of the
existing dynamical models of nucleus-nucleus collisions they are described as
an incoherent  superposition of nucleon-nucleon ones. As we have seen,
to correctly describe the partonic configuration underlying the observed
transverse energy flow
in nucleon-nucleon collisions, mechanisms  beyond conventional collinear
factorization are necessary.
In particular this indicates that the results obtained within minijet
approach based on colinearly factorized QCD, see e.g. the recent review
\cite{E} and references therein, must be taken with care.
 
\begin{center}
{\it Acknowledgements}
\end{center}
I am grateful to Yu.~Dokshitzer, M.~Gyulassy, Yu.~Kovchegov, L.~McLerran
and D.~Soper for useful discussions.

I am also grateful to Theoretical Physics Institute, University of Minnesota,
where this work was started, and Brookhaven National Laboratory, where
it was completed, for kind hospitality. This manuscript has been authored
under Contract No. DE-AC02-98CH10886 with the U.S. Department of Energy.

This work was partially supported by RFBR Grant 00-02-16101

\end{document}